\documentclass[conference,a4paper]{IEEEtran}
\usepackage[noadjust]{cite}
\usepackage[utf8]{inputenc}
\usepackage{amsmath}
\usepackage{amsfonts,amssymb}
\usepackage{array}
\usepackage{graphicx}
\usepackage{booktabs}
\usepackage{afterpage}
\usepackage{float}
\usepackage{amsbsy}
\usepackage{mathtools}
\usepackage{color}
\DeclarePairedDelimiter{\ceil}{\lceil}{\rceil}

\date{\today}
\def\bA{{\mathbf{A}}}  \def\bC{{\mathbf{C}}} \def\bD{{\mathbf{D}}} \def\bE{{\mathbf{E}}}
 \def\bG{{\mathbf{G}}} \def\bH{{\mathbf{H}}}  
  \def\bM{{\mathbf{M}}}  
 \def\bQ{{\mathbf{Q}}}  \def\bS{{\mathbf{S}}} 
\def\bU{{\mathbf{U}}} \def\bV{{\mathbf{V}}} \def\bW{{\mathbf{W}}} \def\bX{{\mathbf{X}}} \def\bY{{\mathbf{Y}}}
\def\bZ{{\mathbf{Z}}}
    
\def\bf{{\mathbf{f}}} \def\bg{{\mathbf{g}}}

\def\C{{\mathbb{C}}}

\begin{document}
\title{Passive RIS vs.\ Hybrid RIS: A Comparative Study on Channel Estimation}
\author{\IEEEauthorblockN{Rafaela Schroeder\IEEEauthorrefmark{1},
Jiguang He\IEEEauthorrefmark{1} and Markku Juntti\IEEEauthorrefmark{1}}
\IEEEauthorblockA{\IEEEauthorrefmark{1}Centre for Wireless Communications, FI-90014, University of Oulu, Finland}
\IEEEauthorblockA{E-mail:\{rafaela.schroeder,jiguang.he,markku.juntti\}@oulu.fi}}
\maketitle

\begin{abstract}
The reconfigurable intelligent surface (RIS) plays an important role in maintaining the connectivity in millimeter wave (mmWave) MIMO systems when the direct channel between the transceivers is blocked. However, it is difficult to acquire the channel state information (CSI), which is essential for the design of RIS phase control matrix and beamforming vectors at the transceivers. In this paper, we compare the channel estimation (CE) performance and achieved spectral efficiency (SE) of the purely passive and hybrid RIS architectures. CE is done via atomic norm minimization (ANM). For the purely passive RIS, we follow a two-stage procedure to sequentially estimate the channel parameters, while for the hybrid RIS we estimate the individual channels at the RIS based on the observations from active RIS elements assuming alternating uplink and downlink training. The simulation results show that the purely passive RIS brings better CE and SE performance compared to the hybrid RIS under the same training overhead. We further consider different setups for the hybrid RIS and study the \textit{tradeoffs} among them. 
\end{abstract}

\section{Introduction}
Millimeter-wave (mmWave) multiple-input multiple-output (MIMO) communication is a key technology for 5G cellular networks and beyond~\cite{alkhateebCEHybridprecod,rappaportmmWave}. Due to the severe path loss over short and medium distances, large antenna arrays at both the transmitter and receiver are necessary to guarantee sufficient received signal power~\cite{heath2016overview,basar2020simris}. The mmWave MIMO channel has different features in comparison with the channel for lower frequencies (e.g., sub-6 GHz), for example, the susceptibility to blockages and a limited number of resolvable paths, leading to the natural channel sparsity. 
Due to the high path loss on the mmWave frequencies, the line-of-sight (LoS) connection is often needed to make the communication range beyond just few meters. However, the LoS path between the transceivers is often unavailable due to frequent blockages~\cite{heath2016overview,basar2020simris}. 

Recently, the reconfigurable intelligent surface (RIS) or intelligent reflecting surface (IRS) has been proposed as an disruptive technology to maintain the connectivity and improve the network coverage in mmWave MIMO communications~\cite{basar2019wireless,huang2019reconfigurable,he2020,nemati2020ris}. A common way of realizing a RIS is a phased array, either uniform linear or planar. By adjusting the phase shifts, RIS can interact with the incident signals and reflect them towards the direction of the receiver regardless of its location, breaking the Snell's law. This means the RIS can work as a smart reflector, which enables the intelligent control of the propagation environment. In the literature, the RIS has been used for achieving different goals, e.g., improved spectral efficiency (SE)~\cite{ActElements, wu2019intelligent}, energy efficiency~\cite{huang2019reconfigurable}, security~\cite{cui2019secure}, and even position accuracy~\cite{wymeersch2019radio,He2020WCNCW}.  

One of the major challenges in mmWave MIMO communication lies in efficient yet accurate channel estimation (CE) to enable the optimal design of precoder/beamformer at both the base station (BS) and mobile station (MS). The problem is even more pronounced in the RIS-aided mmWave MIMO systems, which in general include the BS-RIS channel, the RIS phase control matrix, and the RIS-MS channel. The basic set-up is so called {\emph{passive}{\footnote{For simplicity, we adopt the nomenclature of passive RIS in the literature. However, in practice, the RIS is only nearly passive due to the power consumption for the phase-shift control.}}} RIS with any access to the signal reflected at RIS. This further increases the difficulty of CE. The CE for passive RIS-aided MIMO systems has been studied in~\cite{he2020,CE_He_Zhen-Qing,CE_mirza2019,he2020anm,ardah2020trice}. 
For instance, in~\cite{CE_mirza2019}, Mirza and Ali first estimate the direct BS-MS channel, and then apply the bilinear adaptive
vector approximate message passing (BAdVAMP) algorithm to obtain the channels associated with the RIS. In~\cite{he2020anm}, we proposed a super-resolution two-stage channel parameter estimation procedure via atomic norm minimization (ANM). To be specific, the angles of arrival (AoAs) at the MS and angles of departure (AoDs) at the BS are estimated in the first stage, followed by the estimation of the angle differences associated with the RIS and products of path gains (with one from BS-RIS channel, and the other from RIS-MS channel) in the second stage.

Different from the aforementioned works, Taha \textit{et al.} in~\cite{ActElements} introduce a few \emph{active} elements in the RIS (connected to the radio frequency (RF) down-converter and a consequent baseband processing unit) resulting in a \emph{hybrid RIS architecture}. The target is to simplify the CE process with the cost of increased complexity of RIS implementation. Similarly, in~\cite{oneRFchain}, a few active elements and one RF chains are deployed for CE.

In this paper, we compare the CE via ANM of passive and hybrid RIS configurations for mmWave MIMO system. To the best of our knowledge, this is the first work that compares the CE performance of the hybrid and passive RIS architectures. The comparison between these two different commonly known architectures is inspired by the previous works~\cite{ActElements,oneRFchain}, where the active elements are intentionally proposed for RIS channel estimation. We evaluate their estimation performance in terms of the means square error (MSE) of the channel parameters and gains, the RIS gain, and the approximated spectral efficiency (SE). In the hybrid RIS architecture, the channel parameters of the two individual channels, i.e., between the BS and the RIS as well as the UE and the RIS, are separately estimated at RIS via ANM using \emph{alternating uplink and downlink training}. For the passive RIS, we resort to the two-stage CE procedure, detailed in~\cite{he2020anm}. In the simulations, five different setups are considered for the hybrid RIS. The simulation results show the passive RIS outperforms the hybrid one. Obviously, introducing more RF chains at RIS and increasing training overhead can bring better performance. As a brief summary, we study the CE for two different RIS architectures and find that passive RIS is superior to its hybrid counterpart.

This rest of the paper is organized as follows: Section~\ref{sec: system_model} shows the system model, and the CE is described in Section~\ref{section: CE}, followed by RIS phase control matrix design and the beamformer design in Section~\ref{section: bf}. The performance evaluation, metrics and simulation results are provided in Section~\ref{section: perfomance} with conclusions drawn in Section~\ref{section: conclusions}.

\textit{Notation}: A bold capital letter $\mathbf{A}$ denotes matrix and a lowercase letter $\mathbf{a}$ denotes the column vector, and $()^{\mathsf{H}}$, $()^*$, and $()^{\mathsf{T}}$ denote the Hermitian transpose, conjugate, and transpose, respectively. $\otimes$ denotes the Kronecker product, $\odot$ is the Khatri-Rao product, $\mathrm{vec} (\mathbf{A})$ is the vectorization of $\mathbf{A}$, $\mathrm{diag}(\mathbf{a})$ being a square diagonal matrix with entries of $\mathbf{a}$ on its diagonal, $\|.\|_{\mathrm{F}}$ is the Frobenius norm, $\mathrm{Tr} (\mathbf{A})$ is the sum value of the diagonal elements of $\mathbf{A}$, $\mathbb{T}(\mathbf{A})$ is a Toeplitz matrix,$(\cdot){\dagger}$ denotes the Moore–Penrose inverse, $\mathcal{A}$ is the atomic set, $\mathrm{conv}({\mathcal{A}})$ denotes the convex hull of $\mathcal{A}$ and $\mathbb{E}$ is the expectation operator. 

\section{System Model}
\label{sec: system_model}
The RIS-aided mmWave MIMO system consists of one multi-antenna BS, one multi-element RIS, and one multi-antenna MS illustrated in Fig.~\ref{fig1:RIS}. The number of antennas or reflecting elements on the BS, the RIS, and the MS are denoted by $N_\text{B}$, $N_\text{R}$  and $N_\text{M}$, respectively. We assume the direct BS-MS channel suffers from blockage, which motivates the use of the RIS to assist the data transmission between the BS and the MS via RIS. For simplicity, we adopt an uniform linear array (ULA) for the antennas/elements in this paper, while it is possible to be extended for an uniform planar array (UPA). We consider downlink data transmission and training in the case of passive RIS configuration, while two-way alternating downlink-uplink training is capitalized in the hybrid RIS case. The SE is evaluated for the downlink data transmission after CE. We assume also \emph{uplink-downlink channel reciprocity}.

The propagation channel is composed of two individual channels, e.g., BS-RIS and RIS-MS channels, denoted by $\bH_{\text{B,R}}\in\C^{N_{\text{R}}\times N_{\text{B}}}$ and $\bH_{\text{R,M}}\in\C^{N_{\text{M}}\times N_{\text{R}}}$, respectively. We adopt a block-fading channel, which means that the channel parameters stay constant during the coherence time. By following the geometric channel model, we define $\bH_{\text{B,R}}$ as  
\begin{align}
\label{eq:hbr}
\bH_{\text{B,R}} &= \sum\limits_{l = 1}^{L_{\text{B,R}}} [\boldsymbol{\rho}_{\text{B,R}}]_l \boldsymbol{ \alpha}([\boldsymbol{\phi}_{\text{B,R}}]_l ) \boldsymbol{\alpha}^{\mathsf{H}}([\boldsymbol{\theta}_{\text{B,R}}]_l),\nonumber \\
& = \bA(\boldsymbol{\phi}_{\text{B,R}})\mathrm{diag}(\boldsymbol{\rho}_{\text{B,R}})\bA^{\mathsf{H}}(\boldsymbol{\theta}_{\text{B,R}}),
\end{align}
where $[\boldsymbol{\rho}_{\text{B,R}}]_l$ denotes the $l$th propagation path gain, $\boldsymbol{\alpha}([\boldsymbol{\phi}_{\text{B,R}}]_l )$ and $\boldsymbol{\alpha}([\boldsymbol{\theta}_{\text{B,R}}]_l)$ are the array response vectors as a function of $[\boldsymbol{\phi}_{\text{B,R}}]_l$ and $[\boldsymbol{\theta}_{\text{B,R}}]_l$, where $\alpha$ and $\phi$ are the angles of departure (AoD) and angles of arrival (AoA) of the channel, respectively. $L_{\text{B,R}}$ is the number of distinguishable paths. Considering half-wavelength inter-antenna element spacing, the array response vectors are expressed as $[\boldsymbol{ \alpha}([\boldsymbol{\phi}_{\text{B,R}}]_l )]_{n} =  \exp\{j\pi(n-1)\sin( [\boldsymbol{\phi}_{\text{B,R}}]_l)\}$, for $n=1,\cdots,N_\text{R}$ and $[\boldsymbol{\alpha}([\boldsymbol{\theta}_{\text{B,R}}]_l)]_{n} =  \exp\{j\pi(n-1)\sin( [\boldsymbol{\theta}_{\text{B,R}}]_l)\}$, for $n=1,\cdots,N_\text{B}$ and $j = \sqrt{-1}$. 
The array response matrices $\bA(\boldsymbol{\theta}_{\text{B,R}})$ and $\bA(\boldsymbol{\phi}_{\text{B,R}})$ are defined as 
\begin{equation}
    \bA(\boldsymbol{\theta}_{\text{B,R}}) = [\boldsymbol{\alpha}([\boldsymbol{\theta}_{\text{B,R}}]_{1}), ..., \boldsymbol{\alpha}([\boldsymbol{\phi}_{\text{B,R}}]_{L_{\text{B,R}}})],
\end{equation}
\begin{equation}
    \bA(\boldsymbol{\phi}_{\text{B,R}}) = [ \boldsymbol{\alpha}([\boldsymbol{\phi}_{\text{B,R}}]_{1}), ..., \boldsymbol{\alpha}([\boldsymbol{\phi}_{\text{B,R}}]_{L_{\text{B,R}}})].
\end{equation}

Similarly, the RIS-MS channel $\bH_{\text{R,M}}$ is given by
\begin{align}
\label{eq:hrm}
    \bH_{\text{R,M}} &= \sum\limits_{l = 1}^{L_{\text{R,M}}} [\boldsymbol{\rho}_{\text{R,M}}]_l \boldsymbol {\alpha}([\boldsymbol{\phi}_{\text{R,M}}]_l ) \boldsymbol{\alpha}^{\mathsf{H}}([\boldsymbol{\theta}_{\text{R,M}}]_l ),\nonumber\\
    &=\bA(\boldsymbol{\phi}_{\text{R,M}})\mathrm{diag}(\boldsymbol{\rho}_{\text{R,M}})\bA^{\mathsf{H}}(\boldsymbol{\theta}_{\text{R,M}}) 
\end{align}
where $[\boldsymbol{\rho}_{\text{R,M}}]_l$, $\boldsymbol {\alpha}([\boldsymbol{\phi}_{\text{R,M}}]_l )$, and $\boldsymbol{\alpha}([\boldsymbol{\theta}_{\text{R,M}}]_l)$ are the $l$th propagation path gain, and array response vectors as a function of $[\boldsymbol{\phi}_{\text{R,M}}]_l$ and $[\boldsymbol{\theta}_{\text{R,M}}]_l$, respectively. We define the entire BS-RIS-MS channel based on~\eqref{eq:hbr} and~\eqref{eq:hrm} as
\begin{equation}
\label{eq:hc}
      \bH= \bH_{\text{R,M}}\boldsymbol{\Omega}\bH_{\text{B,R}},
\end{equation}
where $\boldsymbol{\Omega}\in\mathbb{C}^{N_{\text{R}}\times N_{\text{R}}}$ is the diagonal phase control matrix at the RIS with constant-modules entries on the diagonal, i.e, $\left[\mathbf{\Omega}\right]_{k,k}=\exp\left({j\omega_k}\right)$. Also, we define the effective channel $\bG \in \mathbb{C}^{L_{\text{R,M}}\times L_{\text{B,R}}}$ as 
\begin{equation}
\label{eq:g}
    \bG  = \mathrm{diag} (\boldsymbol{\rho}_{\text{R,M}})\bA^ {\mathsf{H}}(\boldsymbol{\theta}_{\text{R,M}}){\boldsymbol{\Omega}}\bA(\boldsymbol{\phi}_{\text{B,R}})\mathrm{diag}(\boldsymbol{\rho}_{\text{B,R}}),     
\end{equation}
considering the propagation path gains, RIS phase control matrix, and the angles associated with the RIS.

\subsection{Passive RIS}
\label{subsection: passive}
When the RIS is purely passive, the RIS has no baseband processing units. Therefore, no observations are available at the RIS, and the CE can only be performed either at the BS (in the uplink) or the MS (in the downlink). For the CE, we aim at extracting the channel parameters in~\eqref{eq:hbr} and ~\eqref{eq:hrm}. With the assumption of block fading channels, we divide the coherence time into two sub-intervals, one for CE and the other for data transmission. Moreover, the CE sub-interval is further divided into $T+1$ blocks. 

We apply the pilot-based training procedure, where BS sends a series of training sequences $\bX_t$ during $t=0,\cdots,T$ to the MS. The signal is reflected at the RIS by $\mathbf{\Omega}_t$, further combined at the MS by $\bW_t$, and received as $\bY_t$. 

In the first stage of CE, i.e., $t =0$, the training matrix is $\mathbf{X}_0 \in \mathbb{C}^{N_{\text{B}}\times N_0}$, the combining matrix is $\mathbf{W}_t \in \mathbb{C}^{N_{\text{M}} \times M_0}$ and the phase control matrix is fixed as $\boldsymbol{\Omega}_0$. This procedure leads to the angles estimation, i.e., AoDs at the BS and AoAs at the MS. In the second stage of CE, our focus is the estimation of the remain channel parameters. We design $\mathbf{X}_t$ and $\mathbf{W}_t$ based on angular estimates (AoDs at the BS and AoAs at the MS) from $\mathbf{Y}_0$, and thus have $\mathbf{X}_t\in \mathbb{C}^{N_{\text{B}}\times L_\text{B,R}}$ and $\mathbf{W}_t \in \mathbb{C}^{N_{\text{M}} \times L_{\text{R,M}}}$.  In other words, in the following blocks, $\mathbf{X}_t$ and $\mathbf{W}_t$ are fixed over all the blocks, while the phase control matrix varies from block to block. The received signals are summarized as
\begin{equation}
     \label{eq_y1}
    \mathbf{Y}_{t}=\mathbf{W}_{t}^{\mathsf{H}}\bH_{\text{R,M}}\boldsymbol{\Omega}_t\bH_{\text{B,R}}\bX_{t}+ \bW_{t}^{\mathsf{H}}\bZ_{t},          
    \mbox{for } t=0,\cdots,T,
\end{equation}
where $\mathbf{Z}_t$ is the additive Gaussian noise with each entry distributed as $\mathcal{CN} (0,\sigma^2)$. 

A two-stage CE approach~\cite{he2020,he2020anm} is applied to sequentially estimate the channel parameters based on $\{\bY_1, \cdots, \bY_T\}$. After the estimation, the MS sends the estimates to the RIS and BS for the design of RIS phase control matrix and beamformer at BS.

\subsection{Hybrid RIS}
\label{subsection: hybrid}
\begin{figure} 
    \centering
    \includegraphics[width=\columnwidth]{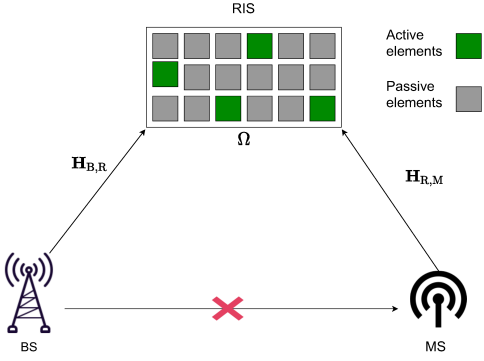}
    \caption{Unlike the passive RIS, a hybrid RIS architecture is a combination of both passive and active elements. In this example, $4$ out of $18$ RIS elements are active, and the rest are passive. }
    \label{fig1:RIS}
\end{figure}

In order to mitigate the difficulty on CE, authors in~\cite{ActElements,oneRFchain} introduce a few active elements to the RIS for collecting signal observations. In such a case, the CE estimation is performed at the RIS with the two individual channels estimated separately. Fig.~\ref{fig1:RIS} shows the hybrid RIS architecture with a combination of both passive and active elements. We offer the example of the hybris RIS in Fig.~\ref{fig1:RIS} with two dimensions for illustration proposes, but our work is focused on the ULA antenna. However, the extension to uniform planar array (UPA) is feasible.

We assume $K$ out of $N_{\text{R}}$ RIS elements are active and $N_{\text{RF}} \leq K$ RF chains are implemented at the RIS.  We consider two-way training in both uplink and downlink. In the downlink, in order to estimate $\mathbf{H}_{\text{B,R}}$, the BS sends training matrix $\bX\in \C^{N_{\text{B}}\times N}$ to the RIS. In the hybrid architecture, the signals are received at the RIS instead of reflecting at RIS and combined at the MS. As a consequence, the received signals model depend only on the number of training from BS. 
The received signals at the RIS are defined as 
\begin{equation}
     \label{eq_y1}
    \mathbf{Y}_{\text{H}}=\bM\bH_{\text{R,M}}\bX +\bM\bZ_{2},  
\end{equation}
where $\mathbf{M}$ is a row-selection matrix containing $K$ rows of a $N_{\text{R}} \times N_{\text{R}}$ identity matrix and $\bZ$ is the additive Gaussian noise with each entry distributed as $\mathcal{CN}(0, \sigma^2)$.


Based on $\bY$, we resort to ANM to extract the channel angular parameter. In the uplink, the same principle is applied to the estimation of parameters in $\mathbf{H}_{\text{R,M}}$ by sending a training matrix with same number of columns as $\bX$ from the MS to the RIS. After the estimation, the RIS feeds back the estimates to the BS and MS for the design of beamformers. 

\section{Channel Estimation}
\label{section: CE}
For the mmWave MIMO system and its evolution with RIS, the CE is efficiently implemented with both on-the-grid compressive sensing (CS) methods such as greedy orthogonal matching pursuit (OMP) and off-the-grid CS methods such as ANM. In general, off-the-grid CS methods can offer higher resolution of parameter estimation compared to its on-the-grid counterparts owing to the continuum of the parameters~\cite{offthegrid}. For this reason, this work focuses on ANM to recover the channel parameters for both the RIS architectures. 
\subsection{Main idea}
In ANM, a signal is represented by using the $\textit{atoms}$ from an $\textit{atomic set}$. The $\textit{atomic set}$ is composed by a finite or infinite number of atoms. The cardinality of the atomic set can be much larger than the dimension of the signal, which results in an infinite number of possibilities to decompose the signal with atoms selected from the pre-defined atomic set. Because of the sparsity, we focus on the decomposition of the signal involving the smallest number of atoms~\cite{harnessing}. 

\subsection{CE via ANM: Hybrid RIS}
\label{subsection: anm}
The hybrid architecture includes the active elements, which are deployed to collect signal observations. It is worth mentioning that the CE is done for all the RIS elements (passive and active). The main difference between the CE procedure for the passive and hybrid RIS is that the active elements enable the explicit channel estimation at the RIS. 

For the hybrid RIS architecture, we apply the ANM for the estimation of the angles, or equivalent spatial frequencies ($\bf = \sin (\boldsymbol{\theta})$ and $\bg = \sin (\boldsymbol{\phi)}$). By replacing the angular parameters with spatial frequencies, we re-write~\eqref{eq:hbr} as 
\begin{equation}
\bH_{\text{B,R}}= \sum\limits_{l = 1}^{L_{\text{B,R}}} [\boldsymbol{\rho}_{\text{B,R}}]_l \boldsymbol{\alpha}([\mathbf{g}]_l ) \boldsymbol{\alpha}^{\mathsf{H}}([\mathbf{f}]_l).
\end{equation}
The spatial frequencies are within $[0,1)$, and the resultant atomic set $\mathcal{A}_{\text{M}}$ is defined as 
\begin{equation}
    \mathcal{A}_{\text{M}} = \left \{ \bQ(f,g): f \in \Big[ 0,1 \Big),  g\in \Big[ 0,1 \Big) \right \},
\end{equation}
where $\bQ(f,g) = \boldsymbol{\alpha}(f) \boldsymbol{\alpha}^{\mathsf{H}}(g)$ is the matrix atom. For matrix $\bH_{\text{B,R}}$, the atomic norm with respect to the atomic set $\mathcal{A}_{\text{M}}$ is formulated as 
\begin{equation}
    \|\bH_{\text{B,R}}\|_{\mathcal{A}_{\text{M}}} = \inf \{q: \bH_{\text{B,R}} \in \mathrm{conv}({\mathcal{A}_{\text{M}}}) \} .
\end{equation}
The equivalent form as a semidefinite programming (SDP) problem is
\begin{equation}
    \nonumber
    \|\mathbf{H}_{\text{B,R}}\|_{\mathcal{A}_{\text{M}}} = \mathrm{inf}_{\{\mathbf{U},\mathbf{V}\}} \Big\{\frac{1}{2 N_\text{B}} \mathrm{Tr}(\mathbb{T}(\mathbf{U})) + \frac{1}{2 N_\text{R}} \mathrm{Tr}(\mathbb{T}(\mathbf{V})) \Big\}
\end{equation}
\begin{center}
    \begin{equation}
        \textrm{s.t}   
        \begin{bmatrix}
        \mathbb{T}(\mathbf{U}) & \mathbf{H}_{\text{B,R}}\\
        \mathbf{H}_{\text{B,R}}^{\mathsf{H}} & \mathbb{T}(\mathbf{V})
    \end{bmatrix}
    \succeq 0.
    \end{equation}
\end{center}
We can recover the angles $\boldsymbol{\theta}_{\text{B,R}}$ and $\boldsymbol{\phi}_{\text{B,R}}$ by addressing the following convex problem
\begin{equation}
    \label{eq:anm_h1}
    \hat{\bH}_{\text{B,R}}= \arg \min_{\bH_{\text{B,R}}} \tau \|\bH_{\text{B,R}}\|_{\mathcal{A}_{\text{M}}} + \frac{1}{2} \| \bM\bH_{\text{B,R}}\mathrm{X} - \bY\|^{2}_{\mathrm{F}}
\end{equation}
where $\tau$ is the regularization parameter set as $\tau \varpropto \sigma\sqrt{N_\text{B} N_\text{R}\log (N_\text{B} N_\text{R}})$. Using the SDP formulation, the problem is further expressed as 
\begin{equation}
            \nonumber
          \hat{\mathbf{H}}_{\text{B,R}} = \arg \min_{\mathbf{{H}},\mathbf{{U}},\mathbf{{V}}} \frac{\tau}{2 N_\text{B}} \mathrm{Tr}(\mathbb{T}(\mathbf{U})) +  \frac{\tau}{2 N_\text{R}} \mathrm{Tr}(\mathbb{T}(\mathbf{V})) 
\end{equation}
\begin{equation}
        \nonumber
        + \frac{1}{2} \|\mathbf{M}\mathbf{H}_{\text{B,R}}\mathbf{X} - \mathbf{Y}\|^{2}_{\mathrm{F}}
\end{equation}
\begin{equation}
    \textrm{s.t} \begin{bmatrix}
    \mathbb{T}(\mathbf{U}) & \mathbf{H}_{\text{B,R}}\\
    \mathbf{H}_{\text{B,R}}^{\mathsf{H}} &  \mathbb{T}(\mathbf{V})\\
    \end{bmatrix}
    \succeq 0,    
\end{equation}
where $\mathbb{T}(\mathbf{U})$ and $\mathbb{T}(\mathbf{V})$ are 2-level Toeplitz matrices. 
The recovery of $\boldsymbol{ \theta}_{\text{B,R}}$ and $\boldsymbol{\phi}_{\text{B,R}}$ is based on the solution of the Toeplitz matrices $\mathbb{T}(\mathbf{U})$ and $\mathbb{T}(\mathbf{V})$, respectively, by applying the ROOTMUSIC algorithm~\cite{MUSIC}. We assume the order information of the angles and the number of paths as priori information. This assumption is necessary due to the loss of order information after the estimation. In practice, the number of paths can be obtained using off-line channel measurement or CS based recovery algorithms~\cite{pathiden}. On the other hand, the order information is used only for the performance evaluation ant it is not a required for the passive and active beamformer design.
The estimation of the path gain vector $\boldsymbol{\rho}_{\text{B,R}}$ is addressed by least squares (LS), which results in
\begin{equation}
    \label{eq_path_gain}
    \hat{\boldsymbol{\rho}}_{\text{B,R}} = 
   \Big[(\mathbf{X}^{\mathsf{T}}\otimes \mathbf{M})\big((\bA^*(\hat{\boldsymbol{\theta}}_{\text{B,R}}) \odot \bA(\hat{\boldsymbol{\phi}}_{\text{B,R}})\big)\Big]^{\dagger}\mathbf{y},
\end{equation}
where $\mathbf{y}$ is defined as $\mathbf{y} = \mathrm{vec}({\bY})$.

The estimation of $\mathbf{H}_{\text{R,M}}$ and parameters therein follows the same procedure assuming similar uplink training. After the estimation of all the channel parameters, we calculate the products of path gains and the angle differences associated with the RIS. These parameters are leveraged to design the RIS phase control matrix, detailed in Section~\ref{section: bf}. The estimated products of path gains $\hat{\boldsymbol{\rho}} \in \C^{L_{\text{B,R}}L_{\text{R,M}} \times 1}$ are defined as 
\begin{equation}
\label{eq_prod_path}
    \hat{\boldsymbol{\rho}} = \hat{\boldsymbol{\rho}}_\text{R,M} \otimes \hat{\boldsymbol{\rho}}_\text{B,R}.
\end{equation}

Therefore, the estimates of the angle differences are functions of the estimates of $\boldsymbol{\phi}_{\text{B,R}}$ and $\boldsymbol{\theta}_{\text{R,M}}$, given by
\begin{align}
\label{eq_ang_diff}
    [\hat{\boldsymbol{\Delta}}]_{lp} &=  \mathrm{asin} \big[ \sin{([\hat{\boldsymbol{\phi}}_{\text{B,R}}]_{l})} - \sin{([\hat{\boldsymbol{\theta}}_{\text{R,M}}]_{p})} \big],\nonumber\\
    &\text{for}\; l = 1,\cdots, L_{\text{B,R}}, p = 1,\cdots, L_{\text{R,M}}. 
\end{align}

Considering the products of path gains~\eqref{eq_prod_path} and the angle differences~\eqref{eq_ang_diff}, we define the $(l,p)$th entry of effective channel~\eqref{eq:g} as 
\begin{align}
    \label{eq:def_g_f}
        [\bG]_{lp}  &= [\boldsymbol{\hat{\rho}}_{\text{R,M}}]_{p}\boldsymbol{\omega}^{\mathsf{T}}\boldsymbol{\alpha}([\hat{\boldsymbol{\Delta}}]_{lp})[\boldsymbol{\hat{\rho}}_{\text{B,R}}]_{l},\nonumber\\
    &\text{for}\; l = 1,\cdots, L_{\text{B,R}}, p = 1,\cdots, L_{\text{R,M}},
\end{align} 
where $\boldsymbol{\omega} = \mathrm{diag}(\boldsymbol{\Omega})$. The vectorization of $\bG$ is $\mathbf{g} = \mathrm{vec}(\bG)$ and the $i$th element of $\mathbf{g}$ is expressed as 
\begin{equation}
   \label{eq:vec_G}
    [\mathbf{g}]_{i} = [\hat{\boldsymbol{\rho}}]_{i}\boldsymbol{\omega} ^{\mathsf{T}}\boldsymbol{\alpha}([\hat{\boldsymbol{\delta}}]_{i}),\;
    \mbox{for }  i=1,\dots,L_{\text{B,R}}L_{\text{R,M}}
\end{equation}
where $\hat{\boldsymbol{\delta}} = \mathrm{vec}(\hat{\boldsymbol{\Delta}})$. 

\subsection{CE via ANM: Passive RIS}
The CE of the passive RIS-aided mmWave MIMO system follows the two-stage procedure using ANM, see~\cite{he2020anm}. We briefly present the CE procedures for this case. In the first stage, we estimate AoDs at the BS ($\boldsymbol{\theta}_{\text{B,R}}$) and AoAs at the MS ($\boldsymbol{\phi}_{\text{R,M}}$) using the multiple measurements vectors (MMV) model. Based on the estimates, we design the beam matrix at the BS and combining matrix at the MS for another round of sounding, where different phase control matrices are considered. In the second stage, we apply the single measurement vector (SMV) model to obtain the remaining channel parameters (angle difference associated with the RIS denoted by $\boldsymbol{\delta}$ and products of path gains $\boldsymbol{\rho}$.) We refer to~\cite{he2020anm} for a more detailed description for this scenario. 

\subsection{Training Overhead}
The training overhead for the hybrid RIS CE is given as
\begin{equation}
    T_{\text{h}} = 2N \Big\lceil \frac{K}{N_{\text{RF}}}\Big\rceil, 
\end{equation}
where $N$ is the number of training beams and the factor $2$ is apply due to the two-way uplink and downlink training procedure. The training overhead for the passive RIS CE in the downlink is 
\begin{equation}
        T_{\text{p}} = N_{0}
        \ceil[\Big]{
        \frac{M_0}{N_{\text{RF},\text{M}}}} + T L_{\text{B,R}}\ceil[\Big]{
        \frac{L_{\text{R,M}}}{N_{\text{RF},\text{M}}}}
\end{equation}
where $N_{0}$ is the number of training beams at the first stage of CE, $M_0$ is the number of columns at the combining matrix at the MS and $N_{\text{RF},\text{M}}$ is the number of RF chains at the MS. 

\section{RIS Phase Control Matrix and Beamforming Design}
\label{section: bf}
The design of the phase control matrix and beamforming vectors follows the same procedure for both the passive and hybrid architecture, which is based on the estimated channel parameters/matrices described in the Section~\ref{section: CE}.
\subsection{RIS Phase Control Matrix}
To design the RIS phase control matrix, one optimization criterion is to maximize the power of the effective channel. The optimal phase control matrix according to this criterion based on the estimates is
\begin{equation}
\label{eq:otimo_omega}
    \boldsymbol{\Omega}^{*}= \arg \max_{\boldsymbol{\Omega}} \|\bG\|^{2}_{\mathrm{F}},
\end{equation}
where $\|\bG\|^{2}_{\mathrm{F}}$ is expressed as
\begin{equation}
\begin{split}\label{eq: norm(g)}
    \|\bG\|^{2}_{\mathrm{F}}&=\|\mathrm{diag}\small(\boldsymbol{\hat{\rho}}_{\text{B,R}}\small)\bA^ {\mathsf{H}}\small(\boldsymbol{\hat{\theta}}_{\text{B,R}}\small){\boldsymbol{\Omega}}\bA\small(\boldsymbol{\hat{\phi}}_{\text{R,M}}\small)\mathrm{diag}\small(\boldsymbol{\hat{\rho}}_{\text{R,M}}\small) \|^{2}_{\mathrm{F}}.\\
\end{split}
\end{equation}
Using~\eqref{eq:g}, we can re-write~\eqref{eq: norm(g)} as 
\begin{equation}
          \|\bG\|^{2}_{\mathrm{F}} = \sum^{L_{\text{B,R}}L_{\text{R,M}}}_{i=1} |[\hat{\boldsymbol{\rho}}]_{i}\boldsymbol{\omega}^{\mathsf{T}}\boldsymbol{\alpha}([\hat{\boldsymbol{\delta}}]_{i})|^2.
\end{equation}
Hence, the optimal $\boldsymbol{\omega}^{*}$ depends on the estimate of the angle difference and the products of path gain with 
\begin{equation}
    \begin{split}
    \label{eq: omega}
    \boldsymbol{\omega}^{*} & = \arg \max_{\boldsymbol{\omega}} \displaystyle \sum^{L_{\text{B,R}}L_{\text{R,M}}}_{i=1} |[\hat{\boldsymbol{\rho}}]_{i}\boldsymbol{\omega} ^{\mathsf{T}}\boldsymbol{\alpha}([\hat{\boldsymbol{\delta}}]_{i})|^2 \\
    &= \arg \max_{\boldsymbol{\omega}} \boldsymbol{\omega}^{\mathsf{T}}\bC \bC^{\mathsf{H}} \boldsymbol{\omega}^{*} \\
    \end{split}    
\end{equation}
where $\bC$ is 
\begin{equation}
\mathbf{C}=\big[\boldsymbol{\alpha}([\boldsymbol{\hat{\delta}}]_{1}),\dots,\boldsymbol{\alpha}([\boldsymbol{\hat{\delta}}]_{L_{\text{B,R}}L_{\text{R,M}}})\big]\mathrm{diag}(\big[[\boldsymbol{\hat{\rho}}]_{1},\dots,[\hat{\boldsymbol{\rho}}]_{L_{\text{B,R}}L_{\text{R,M}}}\big])
\end{equation}
To obtain the optimal $\boldsymbol{\omega}^{*}$, we apply singular value decomposition (SVD) on $\bC\bC^{\mathsf{H}}$, which results in $\mathrm{SVD}(\bC\bC^{\mathsf{H}}) = \bE\bD\bE^{\mathsf{H}}$. 
We select the first column of $\bE$ and then project it to the unit-modulus vector space, resulting in $\boldsymbol{\omega}^{*} = \exp (-j \mathrm{phase}([\bE]_{:,1}))$, where $\mathrm{phase}()$ is the element-wise operation of extracting the phase of the argument.  
\subsection{Beamforming Vectors}
The design of the beamforming vectors at BS and MS, respectively, $\mathbf{f}$ and $\mathbf{w}$, is based on the reconstruction of the entire channel and the design of the phase control matrix. For this end, we resort to the reconstruction of the entire channel $\hat{\mathbf{H}}$, defined as
\begin{equation} 
     \label{eq:est_entire_channel}
    \hat{\bH} = \hat{\bH}_{\text{R,M}}\hat{\boldsymbol{\Omega}}\hat{\bH}_{\text{B,R}}.
\end{equation}
We apply SVD on $\hat{\mathbf{H}}$, resulting in $\hat{\bH} = \bU\bS\bV$ 
where the beamforming vectors are selected as the pair of singular vectors associated with the largest singular value, defined as $\mathbf{f}=[\bV]_{:,1}$ and $\mathbf{w} = [\bU]_{:,1}$.

\section{Performance Evaluation}
\label{section: perfomance}
In this section, we describe 
the parameter setup and the metrics to evaluate the performance. The propagation path gains are defined as $\mathcal{CN} (0,1)$. We assume $N_{\text{B}} = 16$, $N_{\text{R}} = 32$,  $N_{\text{M}}= 16$, $L_{\text{B,R}} = L_{\text{R,M}} = 2$. The SNR is $\frac{1}{\sigma^{2}}$. We perform 2000 trials to average out the results. We consider the following architectures: $\mathrm{(i)}$ Purely passive RIS, $\mathrm{(ii)}$ hybrid RIS, where $K = N_{\text{RF}}$, $\mathrm{(iii)}$ hybrid RIS, where $N_{\text{RF}} < K$, leading to the use of an analog combiner. In case $\mathrm{(iii)}$, the selection matrix is 
replaced by $\bM_{\text{c}}\bM$, where $\bM_{\text{c}}$ is the analog combining matrix with dimensions $N_{\text{RF}}\times K$ and random phases. 

In order to make a fair comparison, we guarantee the same training overhead for all the aforementioned scenarios, e.g., $T_{\text{h}} = T_{\text{p}} = 40$. We also evaluate the performance with different numbers of active elements and RF chains at the RIS. Moreover, we offer the performance of the fully active RIS (i.e., $N_{\text{R}} = K = N_{\text{RF}}$) as an bound for the hybrid RIS. Table 1 summarizes five different parameter setups for the hybrid RIS. 
\begin{table}[!t]
    \centering
    \caption{Parameters setup for the hybrid RIS architecture.}
    \label{tab1:Setups}
    \begin{tabular}{lcccc}
    \toprule
    \textbf  & {$N$} & {$K$} & {$N_{\text{RF}}$} & {$T_{\text{h}}$} \\
   \midrule
        {Setup 1}   &  $20$ & $12$ & $12$ & $40$\\
        {Setup 2}  & $20$ & $16$ & $16$ & $40$\\
        {Setup 3}  & $20$ & $32$ & $32$ & $40$\\
        {Setup 4}  &  $10$ & $12$ & $6$ & $40$\\
        {Setup 5}  &  $10$ & $16$ & $8$ & $40$\\
   \bottomrule
    \end{tabular}
\end{table}

In all the simulations, we use ANM to obtain the estimates thanks to the super-resolution estimation described in~\cite{he2020anm}. In order to guarantee that, we further assume the spatial frequencies are separated by $\left(\frac{4}{N_{\text{B}}}\right)$,  $\left(\frac{4}{N_{\text{R}}}\right)$, $\left(\frac{4}{N_{\text{M}}}\right)$ at the least. 
\subsection{Metrics}
We analyze the performance of the MSE of the estimates of AoDs at BS, AoAs at MS, angle differences and products of path gains. The MSEs for the channel parameters are given as
\begin{equation}
    \text{MSE}(\sin{(\boldsymbol{\phi}_{\text{R},M})}) = \mathbb{E} \Bigg[ \frac{\| \sin{(\boldsymbol{\phi}_{\text{R,M}})} - \sin{(\hat{\boldsymbol{\phi}}_{\text{R,M}})}
    \|^{2}_{2}}{L_{\text{R,M}}} \Bigg],
\end{equation}
\begin{equation}
    \text{MSE}(\sin({\boldsymbol{\delta}})) = \mathbb{E} \Bigg[ \frac{\| \sin{(\boldsymbol{\delta}}) - \sin{(\hat{\boldsymbol{\delta}})\|^{2}_2}}{L_{\text{R,M}}L_{\text{R,M}}} \Bigg],
\end{equation}
\begin{equation}
     \text{MSE}(\boldsymbol{\rho}) = \mathbb{E} \Bigg[ \frac{\| \boldsymbol{\rho} - \hat{\boldsymbol{\rho}}\|^{2}_{\mathrm{2}}}{L_{\text{R,M}}L_{\text{R,M}}} \Bigg].
\end{equation}
We also evaluate the design of RIS phase control matrix via the RIS gain, given by
\begin{equation}
    G_{\text{RIS}}= \frac{\|\bA^{\mathsf{H}}(\boldsymbol{\theta}_{\text{R,M}})
    \boldsymbol{\Omega}^{*}\bA(\boldsymbol{\phi}_{\text{B,R}})\|_\mathrm{F}^2}{N_{\text{R}}^{2}},
\end{equation}
Finally, we express the average SE bound (bits/s/Hz) as
\begin{equation}
    R = \mathbb{E} \Bigg[ \log_{2} \Big(1 + \frac{ |\mathbf{w}^\mathsf{H}\hat{\bH}\mathbf{f}|^{2}}{\sigma^{2}} \Big)\Bigg].
\end{equation}

It is worth mentioning the effect of training overhead is not taken into consideration on the performance evaluation of the average SE bound. However, in all the following simulations, we have a fair comparison that all the training overheads are equal.   

\subsection{Results}
\begin{figure}
    \centering
    \includegraphics[width = \columnwidth]{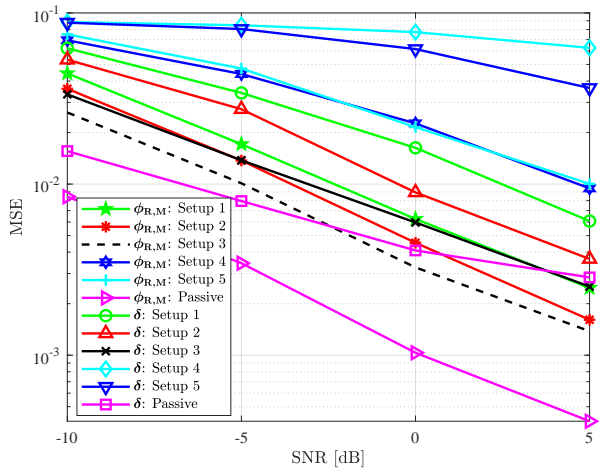}
    \caption{MSE of the channel parameter estimation.}
    \label{fig1:mse_angles}
\end{figure}

The simulation results for the estimation of $\boldsymbol{\phi}_{\text{R,M}}$ and angle difference are shown in Fig.~\ref{fig1:mse_angles}. It is worth mentioning that the estimation of $\boldsymbol{\theta}_{\text{B,R}}$ has similar results as that of $\boldsymbol{\phi}_{\text{R,M}}$, so we omit them here. For the estimation of $\boldsymbol{\phi}_{\text{R,M}}$, the passive RIS outperforms the hybrid RIS. When SNR is $5$ dB, the gap between the passive RIS and setup 3 (lower bound for the hybrid RIS), is the smallest. On the contrary, the gap between the passive RIS and setup 4 is the largest, but setup 4 has the lowest power consumption due to the least number of RF chains in comparison with the other setups. The performance for the estimation of the angle differences is aligned with that of $\boldsymbol{\phi}_{\text{R,M}}$. The passive architecture outperform the overall performance of the hybrid. However, when SNR is $5$ dB, the setup 3 slightly outperform the passive architecture.

\begin{figure}
    \centering
    \includegraphics[width = \columnwidth]{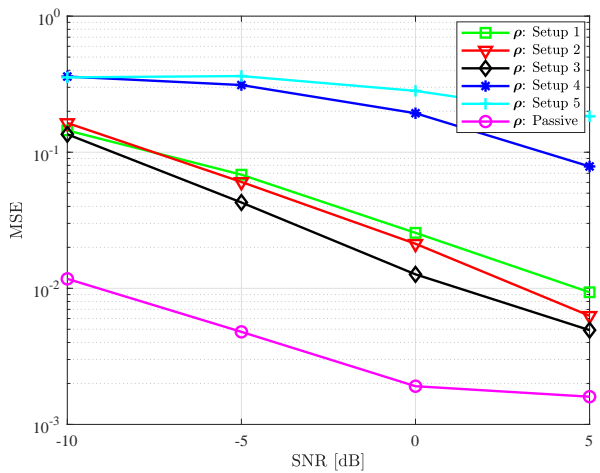}
    \caption{MSE of the products of path gains.}
    \label{fig3: mse path gain}
\end{figure}

Fig.~\ref{fig3: mse path gain} shows the MSE of the products of path gains. The passive RIS also has the best performance in comparison with all the hybrid setups. 
Since we estimate the two individual channels separately, the performance degradation may come from the sharing of the total training overhead and fewer RF chains at the RIS. 

\begin{figure}
    \centering
    \includegraphics[width = \columnwidth]{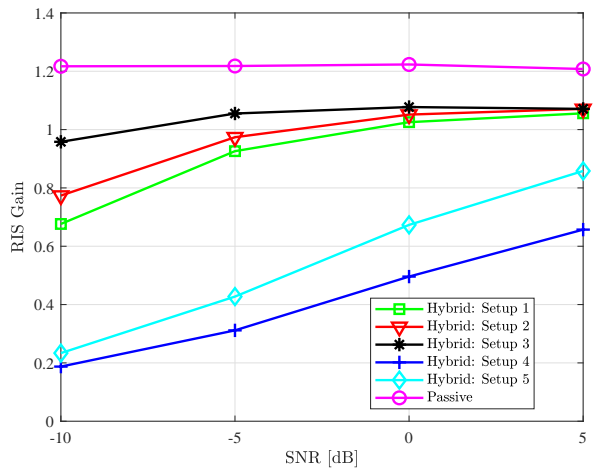}
    \caption{RIS gain.}
    \label{fig2: RIS Gain}
\end{figure}

\begin{figure}
    \centering
    \includegraphics[width = \columnwidth]{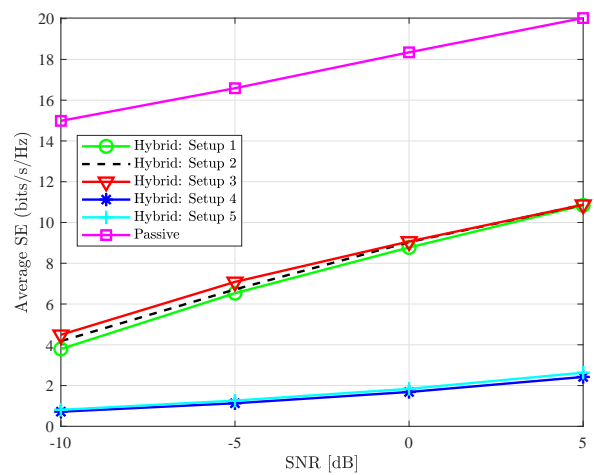}
    \caption{Average SE bound.}
    \label{fig4: capacity}
\end{figure}

For the results of the RIS gain and average SE bound in~Fig.~\ref{fig2: RIS Gain} and ~\ref{fig4: capacity}, setups $1-3$ achieve better results. The estimation of the products of path gains and angle differences are essential to the design of RIS phase control matrix and the beamforming vectors at both the BS and MS. For this reason, the gap in the results of RIS gain and average SE bound between the passive RIS and hybrid RIS setups are even more significant in~Fig.~\ref{fig2: RIS Gain} and ~\ref{fig4: capacity}. Setups $4-5$ have a reduced number of RF chains, which affects the performance in comparison with setups $1-3$, as we can observe from Fig.~\ref{fig4: capacity}. 
\subsection{Discussions}
In our work, the active elements are proposed to collect signal observations. We clarify that the active elements are randomly selected. However, their position does not affect the accuracy of the proposed CE algorithm. It is worth to mentioning that The CE can be improved with a larger number of RF chains and training sequences. For data transmission, the position of the active elements (can adjust both phase and amplitude of incident signals) may yield a better SE, which is left for the future work.

\section{Conclusions}
\label{section: conclusions}
In this paper, we have studied the CE problem for RIS with both passive and hybrid architectures. Our results have shown that passive could bring better performance without additional power consumption. However, the hybrid RIS allows the channel estimation at RIS and decouples the channel estimation problem, and in practice, lower path loss is expected for the two individual channels. Taking this into consideration, the hybrid RIS may bring better performance than passive RIS, which will be left as our future study.  Also, if there exists a backhaul link between the RIS and the MS, the downlink channel estimation performance for the hybrid RIS will be further enhanced if joint processing of received signals (e.g., one at MS and one from RIS) is conducted at the MS. 

\section{Acknowledgements}
This work has been financially supported in part by the Academy of Finland (ROHM project, grant 319485), European Union's Horizon 2020 Framework Programme for Research and Innovation (ARIADNE project, under grant agreement no. 871464), and Academy of Finland 6Genesis Flagship (grant 318927).
\bibliographystyle{IEEEtran}
\bibliography{IEEEabrv,references}

\end{document}